%% file: IEEEPaper.tex
\documentclass[conference]{IEEEtran}

\usepackage{cite}

\ifCLASSINFOpdf
  \usepackage[pdftex]{graphicx}
\usepackage{tikz}
\usepackage{tabularx}

\usepackage{float}
  \graphicspath{{./pdf/}{./jpeg/}}
\else
\fi

\usepackage[cmex10]{amsmath}

\usepackage[ruled,vlined,linesnumbered]{algorithm2e}
\usepackage{color}

\usepackage{xcolor}
\usepackage[thinlines]{easytable}
\usepackage{xcolor}

\begin{document}
%
\title{Towards Adaptive State Consistency \\in Distributed SDN Control Plane}



\author{\IEEEauthorblockN{Ermin Sakic*, Fragkiskos Sardis**, Jochen Guck*** and Wolfgang Kellerer***}
	\IEEEauthorblockA{* \{Technical University of Munich, Siemens AG\}, Munich, Germany, ermin.sakic@\{tum.de, siemens.com\}\\
		** Centre for Telecommunications Research, King's College London, London, UK, fragkiskos.sardis@kcl.ac.uk\\
			*** Technical University of Munich, 
			Munich, Germany, \{guck, wolfgang.kellerer\}@tum.de
	}
}

\usetikzlibrary{arrows.meta}
\usetikzlibrary{backgrounds}
\usetikzlibrary{matrix}

%


\maketitle

\begin{abstract}
State synchronisation in clustered Software Defined Networking controller deployments ensures that all instances of the controller have the same state information in order to provide redundancy. Current implementations of controllers use a strong consistency model, where configuration changes must be synchronised across a number of instances before they are applied on the network infrastructure. For large deployments, this blocking process increases the delay of state synchronisation across cluster members and consequently has a detrimental effect on network operations that require rapid response, such as fast failover and Quality of Service applications. In this paper, we introduce an adaptive consistency model for SDN Controllers that employs concepts of eventual consistency models along with a novel 'cost-based' approach where strict synchronisation is employed for critical operations that affect a large portion of the network resources while less critical changes are periodically propagated across cluster nodes. We use simulation to evaluate our model and demonstrate the potential gains in performance.
\end{abstract}


%
\IEEEpeerreviewmaketitle

\section{Introduction}
\input{introduction}

\section{Problem Definition}
\input{odl_clustering}

\section{Related Work}
\input{relatedwork}

\section{Proposed Solution}
\input{solution}

\section{Implementation}
\input{evaluation}

\section{Conclusion}
\input{conclusion}

\section{Acknowledgement}
{This work has received funding from the European Union's Horizon 2020 research and innovation programme under grant agreement No. 671648 VirtuWind.}

\bibliographystyle{IEEEtran}
\bibliography{IEEEabrv,qos}


\end{document}

%% file: introduction.tex
Software Defined Networking (SDN) is one of the main technologies in 5G networks which enables the logically centralised control of network infrastructures by abstracting the underlying topology and exposing high-level Application Programming Interfaces (APIs) for controlling packet forwarding. The centralized control paradigm however introduces a single point of failure and scalability challenges in large infrastructures such as those found in network service providers and data centres. To address the scalability and resilience issues, controllers can be run in distributed mode, where individual instances participate in load-balancing of network events and provide redundancy in case of controller failures.

For large networks, the scalability and resilience of the SDN controller becomes important as  it poses a single point of failure for the entire network. A single node running the controller may not scale well for hundreds of switches and thousands of concurrent OpenFlow events. Similarly, a single node cannot provide high reliability when it comes to hardware failures. To address this problem, SDN controllers are clustered over multiple nodes, where each instance of the controller is responsible for a number of switches while also providing redundant copies of the other instances' state. When a node fails, another controller instance takes over the failed node's tasks and resumes operation with no downtime. 

In distributed computing, clustering refers to the loose or tight coupling of nodes for purpose of reliability and load balancing. Such systems can be scaled horizontally  by adding nodes to the cluster, however, as more nodes are added, the overhead in state synchronisation between nodes increases. There exist two main models for synchronising state across a cluster. The \emph{strong consistency model} \cite{Panda:2013:CN:2491185.2491186} requires that the distributed state across cluster members is replicated and, following any single state-update at state leader, propagated using mutual consensus to replicas. In contrast, the \emph{eventual consistency model} \cite{Bailis:2013:ECT:2447976.2447992} omits the consensus procedure and guarantees that \emph{at least one delivery} invariant holds. However, the advantage of non-blocking operations comes at the expense of sacrificing the total ordering of state updates and sometimes the system correctness. In eventually consistent systems, the convergence to a single state is determined by two factors: anti-entropy and reconciliation. Anti-entropy ensures that data is synchronised in a timely manner and that the system will not enter a state of complete de-synchronisation between instances \cite{Demers:1987:EAR:41840.41841}. Reconciliation refers to the mechanisms that determine the final system state by resolving conflicting updates from different instances. Typically such conflicts are resolved by the \emph{last-writer-wins} approach, where the most recent change of state is considered final \cite{Vogels:2009:EC:1435417.1435432}.

In this paper, we introduce the concept of runtime adaptation of consistency levels in state synchronisation for an SDN Distributed Control Plane (DCP). A  consistency level is assigned to every resource state accessed by an SDN application (e.g. routing, topology manager). The consistency level is adapted based on the experienced effort of state convergence after a non-synchronisation period has expired; and the inefficiencies resulting from operations with stale state as inputs. We define the application inefficiency as a qualitative distance between the optimal and computed result. The proposed method enables the design of scalable SDN DCPs, since the majority of state updates are executed  as local non-blocking, eventually consistent operations. The methodology of changing the level of controller consistency on-the-fly allows for maintaining a scalable system by sacrificing some controllable amount of result optimality - and thus the blocking overhead of cluster-wide synchronisation.

Inspired by the concept of \emph{demarcation protocol} reservations \cite{barbard1992demarcation}, we introduce \emph{bounded resource credit sets} for safe updates of a shared resource state. By means of operating on a limited set of assigned reservation tokens, manipulation of resources does not result in immediate propagation of a state update, but allows for maintaining the resource bounds invariant and hence correctness property. Contrary to the approach where following a detected conflict consistency levels are adapted, the \emph{bounded resource credits} approach avoids occurence of conflicts and initiates cluster-wide synchronisation only when the controller has depleted its assigned reservation tokens.

The rest of the paper is structured as follows: Section II describes the problem of state synchronisation in distributed SDN controller deployments, Section III investigates the state of art related to state synchronisation in distributed systems and focuses on existing SDN controller implementations, Section IV presents the proposed solution for adaptive state consistency, Section V presents and discusses the performance of the proposed method using simulation and finally, Section VI concludes this paper.

%% file: odl_clustering.tex
In distributed deployments, individual controller instances hold application state necessary to fulfil the requirements of controller applications (e.g. path finding, network firewall, policy handlers). Assuming a partitioned DCP design, we distinguish between \emph{global} and \emph{local} controller decisions. Global decisions necessitate a response to events where an action modifies the configuration of a switch that is outside the controller's administrative domain. Interaction with other controller instances in the DCP is necessary, resulting in latency overhead in controller-to-controller interactions. In state of the art scalable SDN DCPs \cite{Medved, Berde:2014:OTO:2620728.2620744}, a network is typically partitioned into multiple administrative sub-domains.To minimize controller-to-controller synchronisation efforts in the DCP, our model supports transformation of global controller decisions into local ones, by means of assigning all controllers as masters of all switches and granular per-controller planning of switch notification subscriptions for scalability. Hence in our case, a global route configuration that necessitates message passing across the whole DCP in current controllers \cite{Medved, Berde:2014:OTO:2620728.2620744}, can be applied by a single controller to all switches on path, since the administrative domain of the controller can stretch across the whole network. The related OpenFlow-based controller role configuration is explained in more detail in Subsection \ref{sec:model}. 

\vspace{5mm}
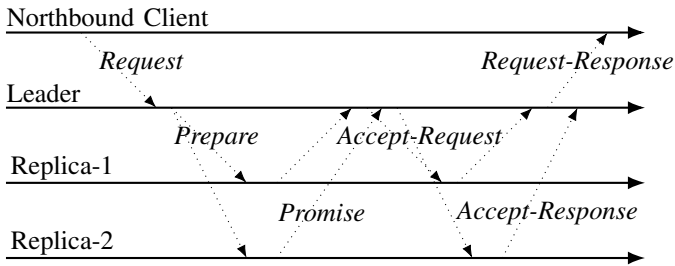
\begin{figure} [H]
	\begin{tikzpicture}
	
	\draw[thick, ->, -Latex]  (0,1) -- (8.5,1) ; 
	\draw[thick, ->, -Latex]  (0,2) -- (8.5,2) ;
	\draw[thick, ->, -Latex]  (0,3) -- (8.5,3) ;
	\draw[thick, ->, -Latex]  (0,4) -- (8.5,4) ;
	
	\draw[dotted, ->, -Latex] (1,4) --  (2,3); 
	
	\draw[dotted, ->, -Latex] (2.2,3) --  (3.2,2); 
	\draw[dotted, ->, -Latex] (2.2,3) --  (3.2,1); 
	
	\draw[dotted, ->, -Latex] (3.6,2) --  (4.6,3); 
	\draw[dotted, ->, -Latex] (3.6,1) --  (5,3); 
	
	\draw[dotted, ->, -Latex] (4.8,3) --  (5.8,2); 
	\draw[dotted, ->, -Latex] (5.2,3) --  (6.2,1); 
	
	\draw[dotted, ->, -Latex] (6,2) --  (7,3); 
	\draw[dotted, ->, -Latex] (6.6,1) --  (7.6,3); 
	
	\draw[dotted, ->, -Latex] (7.2,3) --  (8,4); 
	
	\node [overlay, align=left] at (1.35,4.2) {Northbound Client};
	\node [overlay, align=left] at (0.5,3.2) {Leader};
	\node [overlay, align=left] at (0.75,2.2) {Replica-1};
	\node [overlay, align=left] at (0.75,1.2) {Replica-2};
	
	\node [overlay, align=left] at (1.8,3.6) {\it{Request}};
	\node [overlay, align=left] at (2.8,2.6) {\it{Prepare}};
	\node [overlay, align=left] at (4.2,1.6) {\it{Promise}};
	\node [overlay, align=left] at (5.5,2.6) {\it{Accept-Request}};
	\node [overlay, align=left] at (7.2,1.6) {\it{Accept-Response}};
	\node [overlay, align=left] at (7.6,3.6) {\it{Request-Response}};
	
	\end{tikzpicture}
	\caption{Paxos workflow where leader requires confirmation only from cluster majority to progress the state. Notice how Replica-2 delays its response. } 
	\label{fig:paxos}
\end{figure}

Strong consistency systems always implement some consensus algorithm to enable conflict-free distribution of state updates. Paxos \cite{paxos}, a popular decentralized consensus algorithm, proposes a four-delay state update method encompassing \emph{Prepare-Request}, \emph{Promise}, \emph{Accept-Request} and \emph{Accept-Response} delays, as depicted in Figure \ref{fig:paxos}.  The \emph{Prepare} phase, and hence two propagation delays, may be skipped if the \emph{Proposer} instance knows that it is the only one to suggest a value. In large-scale SDN deployments, the amount of incoming controller requests can reach up to 11 million requests per second \cite{dcn}. In worst case, every request may necessitate a consensus run per state change, thus preventing fast network reconfigurations and introducing a bottleneck on control plane, and ultimately data-plane.  Although slight variations of Paxos, including the recent RAFT \cite{184040} consensus algorithm, were proposed in literature \cite{184040, ho2016fast, du2014clock}, the general concept and signaling overhead of the algorithm is unchanged.

Recently published industrial SDN use-cases introduce new requirements on global QoS-aware route establishment across Wide Area Networks (WAN) \cite{petropoulossoftware, bianco2016role} and locally administered networks \cite{7517394}. Petropoulos \emph{et al.} \cite{petropoulossoftware} describe the requirements of critical infrastructure operators for on-demand network service establishment in a Software Defined-WAN and for the interconnection of a large number of IoT (Internet of Things) devices over network service provider's infrastructure. Their wind-power use-case assumes QoS guarantees for individual IoT-device-to-Cloud application flows. Coupled with the ever-growing number of IoT devices and need for dynamic resource (de-)allocation for globally computed QoS-enabling paths, current SDN controller solutions will not be able to provide the necessary degree of scalability in global configuration. Some 5G use-cases \cite{5gic} introduce the requirements of connection setup times of $<$15-30ms for low-latency services in converged backbone for  arbitrary numbers of end-hosts. Initial performance measurements of an SDN- and OpenDaylight-enabled DCP in the test network of Telecom Italia \cite{bianco2016role} show that the requirement of low setup time cannot be met in most scenarios. The identified main cause of delay in end-to-end path establishment lies in the controller-to-controller interactions, required in order to reach consensus for distributed path establishment. Our approach can minimize this response handling time in critical path by means of adaptively lowering the frequency of controller-to-controller interactions, hence enabling a larger throughput of new connection admissions than possible with current strong consistent DCP models.

Alternative approaches to the strong consistency model assume an eventual consistent state synchronisation where changes made to a controller instance get propagated over time across the cluster, thus solving the issue of blocking during synchronisation period. This allows the active instance to apply changes immediately and synchronise its state over time. However, it may also lead to a scenario where, upon a node's failure, the current state of a sub-domain is lost. This can have a detrimental effect on the entire network as the consistency of initial state determines the quality of output delivered by the controller's decisions. For example, a path computation application which identifies a globally optimal path on the network with regards to currently utilized resources and a given set of constraints (e.g. on delay, bandwidth etc.), may produce a suboptimal result due to stale information on the state of reservations. We consider the amount of observed result suboptimality as an input for our autonomous consistency level adaptation algorithm. This online algorithm outputs the consistency level required in order to provide an \emph{exactly sufficient} amount of experienced result-correctness.

Summarized, this paper investigates the relation between overhead minimization in SDN DCP and associated system correctness. In particular, we investigate the trade-off between lowered response-time delays and inefficiencies resulting from operating with stale data. Can strict requirements on low setup times be supported for different topologies and traffic patterns, while ensuring a sufficient degree of system correctness at all times? We conclude that a strong consistent DCP introduces critical overhead in controller-to-controller synchronisation \cite{bianco2016role}, while an eventually consistent DCP provides no correctness guarantees whatsoever \cite{Panda:2013:CN:2491185.2491186}. Hence we introduce an alternative adaptive consistency model that provides instantaneous response for majority of requests, while bounding the observed correctness of result to a tunable threshold.

%% file: relatedwork.tex

In this section we investigate the state-of-art in distributed systems state synchronisation. Special focus is placed on clustering mechanisms in existing SDN controller implementations and their advantages and disadvantages in large-scale deployments.
	
Botelho \emph{et al.} \cite{Botelho:2013:FCF:2570448.2570470} investigate the performance of a strongly-consistent replicated data store implementation BFT-SMaRT \cite{Bessani2014StateMR} in Floodlight. While the results in a four-node cluster show promising transaction throughput for data store requests made for simple host-port mappings, significantly lowered data-store performance was measured for more complex load balancing and device management operations. The enforcement delays and blocking times in data-plane were not considered in this study. Furthermore, the evaluated network consisted of a single OpenFlow switch, deployed in an out-of-band control channel. A more realistic in-band control network would cause higher variance in request inter-arrival times, higher data-plane configuration latencies and possibly data-plane bottlenecks in case all requests were destined for a centralized cluster leader.

In terms of SDN controller solutions, HyperFlow \cite{tootoonchian2010hyperflow} selectively synchronises the network state among controller replicas via an eventually consistent publish/subscribe system based on WheelFS \cite{stribling2009flexible}. OpenDaylight (ODL) \cite{Medved} and DISCO \cite{DBLP:journals/corr/PhemiusBL13} rely on strong synchronisation between all controller instances, where any changes made to a particular instance have to be synchronised with a number of other instances in the cluster before they can be applied to the network. In ODL Clustering, a single controller instance is the leader of any state shard at any point in time and only the leader is allowed to initialize changes to its state. Following a state modification, the leader propagates the update to follower-controller replicas that also hold the shard. The leader, together with its followers make up a strong consistency cluster. ODL Clustering implements RAFT \cite{184040}, a consensus algorithm which extends Paxos \cite{paxos} with membership change and leader election mechanisms, but is equivalent in terms of fault-tolerance and performance. Inside the RAFT cluster of size $N$, each data store change is initiated by the leader, and propagated to at least $N/2$ followers. The followers must acknowledge the state update before the leader can continue processing further state changes. This blocking period may take an arbitrary amount of time depending on the placement of leader and followers, propagation delays in the network, processing delays and size of quorum. The requirement that at least $N/2+1$ cluster participants reflect the leader’s state update is a minimum requirement for overlapping reads and writes in a reliable cluster that tolerates $\lfloor{N/2}\rfloor$ failures \cite{paxos}. More stringent deployments may require acknowledgement by a higher number of followers, hence causing additional overhead.

ONOS \cite{Berde:2014:OTO:2620728.2620744} and ONIX \cite{Koponen:2010:ODC:1924943.1924968} expose an alternative and more recent controller design that tries to solve the issue of scalability by providing the APIs for selection of either strong or eventual consistency mode for its distributed state primitives. Thus, applications which can operate correctly without strictly consistent state updates synchronise in eventually consistent manner. However, the active state consistency model does not change at run-time and must be hard-coded in the SDN application without knowing the exact constraints of the network it will be deployed in. The application’s designer is unaware if the application might run correctly even if its state was eventually synchronised – which is the case when the probability of a state-conflict is very small. This may lead to overly-pessimistic estimations of an application's requirements in the deployed domain. For example, routing applications may tolerate suboptimality if maximized network utilization is not a concern. 

Levin \emph{et al.} \cite{Levin:2012:LCS:2342441.2342443}, show that distributed network functions, such as load-balancers, can work around eventual consistency and still deliver performance sufficient for production deployments. Yu \emph{et al.} \cite{Yu:2000:DEC:1251229.1251250} introduce a \emph{continuous consistency model} for geo-replicated services, where application designers can bound the maximum distance between the local data state and final consistent state. In their model, distance in actual and stale state view is parametrized by the numerical and order error and state staleness. In context of SDN, Panda \emph{et al.} \cite{Panda:2013:CN:2491185.2491186} argue that linearisability is likely an unnecessary property for ensuring correct application of most network policies, as the investigated policies often have simple correctness conditions. Furthermore, the authors state that determining a consistency model that is \emph{necessary} and \emph{sufficient} for network policies is an important research problem.  


%% file: solution.tex

In the following section, we introduce our adaptive consistency model for SDN controllers, where state synchronization occurs according to performance and consistency constraints set by the application at runtime. We make use of triggers that allow for dynamic switching of consistency level on a per-state-fragment basis, based on a defined local threshold. This threshold could, for example, be the allowed observed suboptimality of path reservations, based on consistency of the resource reservation state of accepted paths. In this case, the suboptimality of a result has a source in the fact that concurrently executed path reservations were not propagated to the instance that established the suboptimal path. However, the trade-offs of suboptimality and scalability might be tolerable in systems where request throughput and response time are highly-valued properties. 

Our model envisions an eventually consistent synchronisation approach in an SDN DCP, where synchronisation credits for state modifications are assigned to controller replicas. The credit-based approach bounds the data staleness which may arise as a consequence of concurrent and non-synchronised modification of states in cluster members. Consistency levels control the frequency of DCP-wide state-updates and eventual reconciliation of state-conflicts. Tight consistency levels lead to more frequent synchronisation, hence causing more control plane overhead compared to relaxed consistency levels, where synchronisation overhead is minimized but the probability of occurring state-conflicts is raised. Our cost-aware algorithm is used to identify the balance in between the two key performance indicators; the \emph{synchronisation overhead} and \emph{system correctness} of DCP, by adaptively modifying the consistency levels based on application-specified correctness thresholds. We also describe conflict detection and remediation mechanisms that serve the purpose of deterministic convergence to a single stable state on conflicting controller replicas.

\subsection{System Model}
\label{sec:model}
The OpenFlow-based DCP is modeled as a cluster of $N$ controllers with each switch in the data plane configured to register with all controllers in the administrative domain in \emph{OFPCR\_ROLE\_EQUAL} mode. In this mode, the controller has full access to the switch and is equal to other controllers of the same role. Controllers react to external events (northbound requests, \emph{PACKET\_IN}s and other OpenFlow notifications) locally. While northbound requests are directed and hence handled by a single controller instance, the invariant of the \emph{exactly-once} response needs to hold for asynchronous data plane events (e.g. \emph{PACKET\_IN}s) as well. By default, all controllers in \emph{OFPCR\_ROLE\_EQUAL} role receive all switch asynchronous messages. Hence, \emph{OFPCR\_EQUAL} mode in OpenFlow must be combined using per-controller \emph{Asynchonous Configuration} \cite{version1} to expose notifications (e.g. \emph{PACKET\_IN}s) to a subset of controllers. The choice of the exact controller that handles a switch request is a controller assignment problem \cite{Wang2016DynamicSC} and is out of scope in this paper. Alternatively, the switch can broadcast the request to all controllers in \emph{OFPCR\_EQUAL} role. Following a successful execution of response, the \emph{exactly-once} execution invariant can be guaranteed by broadcasting a proprietary notification to the assigned controller set \cite{mantas2016consistent}.
 
 
To circumvent the need for a strictly consistent cluster synchronisation, a resource state $S$ in our system is associated with a maximum synchronisation credit amount  $C^{S}_{total}$. Every credit value $C_S$ represents a smallest non-divisible element of resource $S$, allocated to an SDN controller instance $K_N$ for concurrent and non-synchronised modification of a shared resource state $S$. By bounding the amount of concurrent modifications for resource $S$ per controller, distance in value between the local, concurrently modified, data state $S$ and its final consistent state (after synchronisation) can be controlled. The total sum of synchronisation credits $C^{S}_{total}$ represents the maximum number of concurrent modifications to state $S$ during a single non-synchronisation cycle. We distribute the total synchronisation credits across $N$ controller so that:
 
\begin{equation}
	C^{S}_{total} = \sum_{K_N=0}^N C^{S}_{K_N}
\end{equation}
Resource representation of $S$ may encompass both physical or virtual network resources - e.g. bandwidth or flow table elements available for reservation. Depending on the required granularity of state management, limitations for bounded number of modifications may be configured either per state or per operation which affects multiple states. Hence we distinguish between \emph{resource credits} $C^{S_{}}_{total}$ for state \emph{S}, and \emph{execution credits} $C^{{Op}}_{total}$ for operation \emph{Op} that modifies state \emph{S}.

Isolation of state modifications per-controller allows for concurrent and unsynchronised access to state. In Figure \ref{fig:topologymodel}, the resource credit $C^{S_{BW}}_{total}$ for bandwidth resource $S_{BW} =  S_{BW}^{{S1 \rightarrow S2}}$ for edge $S1 \rightarrow S2$, and the execution credit $C^{{add-flow}}_{total}$ for operation \emph{add-flow} that operates on state $S_{BW}^{{S1 \rightarrow S2}}$, are distributed across the controller instances.

 \newcommand*\ab{.4}
 \tikzset{
 	net ctl/.style = {circle, minimum width=2*\ab cm, inner sep=0pt, outer sep=0pt, fill={rgb:orange,1;yellow,2;pink,5}, line width=0.3mm, draw},
 	net sw/.style = {rounded corners,minimum width=2*\ab cm, minimum height=1.5*\ab cm, inner sep=0pt, outer sep=10pt, line width=0.3mm, draw,fill={rgb:black,1;white,4}},
 	net root node/.style = {net node, minimum width=3*\ab cm},
 	net connect/.style = {line width=0.3mm, draw, draw=blue!50!cyan!25!black},
 	net connectred/.style = {line width=1.5pt, draw=red},
 }
 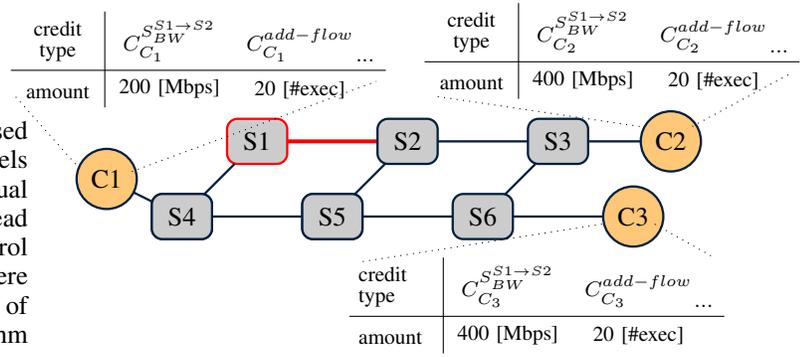
\begin{figure}

 	\begin{tikzpicture}
 	\path [net connect] (1,1) -- (0,1.5) node [net ctl] {C1};
 	\path [net connect] (6,2) -- (7.5,2) node [net ctl] {C2};
 	\path [net connect] (5,1) -- (7,1) node [net ctl] {C3};
 
  	\path [net connectred] (4,2) -- (2,2) node [net sw] {S1};
  	\path [net connect] (6,2) -- (4,2) node [net sw] {S2};
  	\path [net connect] (6,2) -- (6,2) node [net sw] {S3};
  	
  	\path [net connect] (1.7,1.7) -- (1,1) node [net sw] {S4};
  	\path [net connect] (3,1) -- (1,1) node [net sw] {S4};
  	\path [net connect] (3.7,1.7) -- (3,1) node [net sw] {S5};
  	\path [net connect] (5,1) -- (3,1) node [net sw] {S5};
  	\path [net connect] (5.7,1.7) -- (5,1) node [net sw] {S6};
  	
 	\end{tikzpicture}

 	\begin{tikzpicture}[overlay]	
 	\draw[dotted] (0,1.5) --  (-0.8, 2.4); 
 	\draw[dotted] (0.75,1.4) --  (4, 2.6); 
 
  	\draw[dotted] (7.5,2) --  (4.8,2.4); 
  	\draw[dotted] (8.2,2) --  (9.5,2.7); 
  	
  	\draw[dotted] (7,0.7) --  (3.8, 0.2); 
  	\draw[dotted] (7.7,0.7) --  (8.5, 0.25); 
  	 	
 		\matrix(dict)  at (1.5,2.5)  [matrix of nodes, overlay,
 		nodes={align=center,text width=0.3cm},
 		row 1/.style={overlay,font=\fontsize{8}{8}\selectfont,anchor=south},
 		column 1/.style={font=\fontsize{8}{7}\selectfont,nodes={text width=1cm,align=center}},
 		row 2/.style={font=\fontsize{8}{8}\selectfont,anchor=south},
 		column 2/.style={font=\fontsize{6}{6}\selectfont,nodes={text width=1.5cm,align=center}},
 		row 3/.style={font=\fontsize{6}{6}\selectfont,anchor=south},
 		column 3/.style={font=\fontsize{6}{6}\selectfont,nodes={text width=1.5cm,align=center}},
 		]{
 			credit type & $C^{S_{BW}^{{S1 \rightarrow S2}}}_{C_1}$ & $C^{{add-flow}}_{C_1}$ & $...$\\
 			amount & 200 [Mbps] & 20 [\#exec]\\
 		};
 		\draw(dict-1-1.south west)--(dict-1-4.south east);
 		\draw(dict-1-1.north east)--(dict-2-1.south east);
 	\end{tikzpicture}
 	
 	\begin{tikzpicture}[overlay]	
 	\matrix(dict)  at (7,3.0)  [matrix of nodes, overlay,
 	nodes={align=center,text width=0.3cm},
 	row 1/.style={overlay,font=\fontsize{8}{8}\selectfont,anchor=south},
 	column 1/.style={font=\fontsize{8}{7}\selectfont,nodes={text width=1cm,align=center}},
 	row 2/.style={font=\fontsize{8}{8}\selectfont,anchor=south},
 	column 2/.style={font=\fontsize{6}{6}\selectfont,nodes={text width=1.5cm,align=center}},
 	row 3/.style={font=\fontsize{6}{6}\selectfont,anchor=south},
 	column 3/.style={font=\fontsize{6}{6}\selectfont,nodes={text width=1.5cm,align=center}},
 	]{
 		credit type & $C^{S_{BW}^{{S1 \rightarrow S2}}}_{C_2}$ & $C^{{add-flow}}_{C_2}$ & $...$\\
 		amount & 400 [Mbps] & 20  [\#exec]\\
 	};
 	\draw(dict-1-1.south west)--(dict-1-4.south east);
 	\draw(dict-1-1.north east)--(dict-2-1.south east);
 	
 	\end{tikzpicture}
 	
 	\begin{tikzpicture}[overlay]	
 	\matrix(dict)  at (6,0)  [matrix of nodes, overlay,
 	nodes={align=center,text width=0.3cm},
 	row 1/.style={overlay,font=\fontsize{8}{8}\selectfont,align = center, anchor=south},
 	column 1/.style={font=\fontsize{8}{7}\selectfont,nodes={text width=1cm,align=left}},
 	row 2/.style={font=\fontsize{8}{8}\selectfont,anchor=south},
 	column 2/.style={font=\fontsize{6}{6}\selectfont,nodes={text width=1.5cm,align=center}},
 	row 3/.style={font=\fontsize{6}{6}\selectfont,anchor=south},
 	column 3/.style={font=\fontsize{6}{6}\selectfont,nodes={text width=1.5cm,align=center}},
 	]{
 		credit type & $C^{S_{BW}^{{S1 \rightarrow S2}}}_{C_3}$ & $C^{{add-flow}}_{C_3}$ & $...$\\
 		amount & 400 [Mbps] & 20  [\#exec]\\
 	};
 	\draw(dict-1-1.south west)--(dict-1-4.south east);
 	\draw(dict-1-1.north east)--(dict-2-1.south east);
 	
 	\end{tikzpicture}
 	\caption{An exemplary assignment of synchronisation credits to controllers in DCP. As \emph{add-flow} operation ultimately modifies the reserved resource amount $S_{BW}^{{S1 \rightarrow S2}}$, notice that both the execution credits $C^{{add-flow}}$, and granular resource credits $C^{S_{BW}^{{S1 \rightarrow S2}}}$ limit the maximum duration of non-synchronisation period for the bandwidth resource $S_{BW}^{{S1 \rightarrow S2}}$. Hence, it is expected that for state \emph{S} the controller tracks a single type of synchronisation credit - $C^S$, at granularity of state $S$; or $C_{Op}$, at granularity of operation $Op$ that modifies $S$.} 
 	\label{fig:topologymodel}
 \end{figure}
 
Controller $K_N$ modifies the resource state in a manner where each update is handled locally for some non-synchronisation period $T^{S}$, without cluster-wide synchronisation of state. $T^{S}$ is thus the time period elapsed in-between cluster-wide synchronisations of controllers' views of state $S$.

In some cases, the strategy of concurrent modifications of a state $S$ may result in global inefficiencies and suboptimality of a result. The \emph{quality of a result} is correlated with the staleness of the input state  $S$. To limit the effect of asynchronous access to a state, we introduce the notion of \emph{consistency levels}. The choice of a consistency level $CL_S$ for the state $S$, defines the maximum duration of non-synchronisation period $T^{S}_{max}$. In our model, the actual elapsed non-synchronisation period $T^{S}_i$ is not the same for all synchronisation cycles $i$ and may vary based on the occurrence frequency of synchronisation triggers. Observed system KPIs, such as the logged network utilization or encountered number of state synchronisation conflicts, are used as triggers for cluster-wide synchronisation procedure. The triggers eventually lead to adaptation of $CL_S$ and are explored further in the text. The result of consistency model adaptation is the new active consistency level. In addition to $T^{S}_{max}$, the active consistency level also governs the maximum number of locally executed, non-synchronised state updates by tightening or relaxing the resource/execution synchronisation credit $C^{S_{}}_{K_N}$ / $C^{Op}_{K_N}$. The synchronisation credit is associated with a particular resource for which the adaptation is executed. 

For example, SDN applications utilizing the edge-cost state for purposes of routing on a network graph may tolerate low consistency for edge-cost values. By decoupling the synchronisation and routing operations, scalability of routing execution is raised at the expense of result optimality. On the other hand, the same routing applications might require strong consistent graph state updates. No routing artefacts such as loop occurrences or blackholes may happen when the topology state is updated, hence topology state modifications must be serialized. Synchronisation of topology updates is in most networks not a scalability constraint since, compared to frequency of routing requests, graph updates are rare.

\subsection{Allocation of Synchronisation Credits}
Manipulation of locally-assigned resources allows for reservation and freeing of the resource without a cluster-wide state synchronisation required, as long the reservations are conducted within the assigned credit boundaries. We distinguish between two types of synchronisation credits: \emph{Execution credit sets} associated with the amount of allowed local operation executions; and \emph{resource credit sets} associated with the actual physical or virtual network resources.
 
\textbf{Bounded execution credit sets.} 
\label{exec-credit-set}
Credit set $C^{{Op}}_{K_N}$ represents the total number of executions of operation $Op$ that may run locally on controller $K_N$, without the cluster-wide synchronisation of states modified by the operation. Different to \emph{demarcation protocol} approach described below, the number of allowed remaining executions drops monotonically with time. Following a depletion of execution credits, cluster-wide synchronisation leads to view convergence for all states modified by the operation $Op$. The credits are then refreshed on all controller replicas.

\textbf{Bounded resource credit sets.} A global resource $S$ is divided into equally sized \emph{N} resource partitions, which are placed on and handled separately by \emph{N} controllers. Manipulation of locally-assigned resources allows for reservation and freeing of the resource, with no synchronisation required in between the holders of the resource, as long as the reservations are conducted within the assigned boundaries. For every resource $S$, each controller $K_N$ is assigned the local resource modification boundaries $inf(S) = 0$  and $sup(S) = C^{S_{}}_{K_N}$. Initially, the controller-internal reservation variable $reserved (S) = 0$, since no resources are reserved at controller initialization time.

This approach assumes that every controller maintains $inf(S)$ and $sup(S)$, as well as the $reserved(S)$ variables for any divisible resource $S$, taking into account all possible outcomes of isolated transactions. Resource $S$ may be manipulated by more than a single application in SDN controller. The resource boundaries are enforced across all controller applications operating on state $S$. Defined manipulations on resource $S$ include increment and decrement operations. Before executing the reservation, the resulting variable $reserved (S)$ is validated against the assigned $inf(S)$ and $sup(S)$, which it may not exceed. This step is the resource access control.  By calling the $decr(S, n)$ operation, the replica consumes $n$ resource tokens. Analogously, by calling $incr(S, n)$, a replica produces $n$ tokens. Tokens are added or subtracted from the current value of $reserved (S)$, following a successful commit of transaction associated with the reservation. A successful commit updates $reserved (S)$ to currently utilized resource amount and unexpected abort of transactions triggers a cluster-wide conflict reconciliation.  In cases wherein no tokens are available:
\begin{equation}
sup (S) - reserved (S) \le inf (S)
\end{equation} tokens may be requested from other resource controllers. If no resource credits are transferred to requester controller, the reservation fails.

All resources whose reservations may be represented as counter objects with fixed ranges can also be modeled as resource credits, some examples being flow table sizes, meter IDs, IP address pools for DHCP, per-port/queue bandwidth shares etc.

\subsection{Adaptation of Consistency Levels}
Assuming unbounded resource credit sets, the choice of a consistency level governs the maximum non-synchronisation period $T^{S}_{max}$ for state $S$. In case of an execution credit set, it governs both $T^{{S}}_{max}$ (where $S$ may be modified by operation $Op$) and the number of allowed isolated executions $C^{{Op}}_{K_N}$ for operation \emph{Op} in SDN controller $K_N$. Consistency level change is state-/operation-specific, hence $C^{S_A}_{K_N} \neq C^{S_B}_{K_N}$ for resources $S_A$ and $S_B$, and $C^{Op1}_{K_N} \neq C^{Op2}_{K_N}$ for operations $Op1$ and $Op2$,.

Following triggers lead to tightened $CL_S$, which, depending on type of resource, result in shorter duration of $T^S_{max}$ and a lower-than-current number of allowed isolated state modifications $C^{S}_{K_N}$ / operation executions $C^{{Op}}_{K_N}$:
\begin{itemize}
	\item \emph{Cost of result suboptimality} of an eventually-consistent execution of an operation is above a specific threshold. Serialized historical information about the requests is required to derive the actual costs of execution. In case path finding computes paths which are considerably suboptimal than would be the case with strong synchronisation, $CL_S$ is tuned to a stricter level.
	
	\item \emph{Cost of state-update conflicts} is above a specified threshold - if all controllers are able to run operations that modify the shared resource set $S$, following a successful conflict detection (e.g. same state is modified concurrently), controllers may raise the $CL_S$ to a stricter level. Conflict resolution strategy must be deployed in order to reconcile the diverged state into a consistent state across the cluster.
	
	\item \emph{Setup-failures} wherein the callback associated with resource reservation does not result in a successful configuration of an external network device. Device is able to detect and notify false configuration, and controller must deduce a configuration conflict. Historical information of steps leading to setup failures is required for conflict detection and state reconciliation. 
	
\end{itemize}
Following triggers lead to relaxation of $CL_S$ , which, depending on type of resource, result in longer $T^S_{max}$ and a higher-than-current number of allowed isolated state modifications $C^{S}_{K_N}$ / operation executions $C^{{Op}}_{K_N}$:
\begin{itemize}
	\item  \emph{Cost of result suboptimality} of an eventually-consistent execution of an operation is below a specific threshold - the result could on average be close-to-optimal even in case of a more relaxed consistency model. 
	\item \emph{Cost of consistency-related conflicts} is below a specific threshold and the observed probability of incurred state-update-related conflicts is small.
\end{itemize}

\subsection{State Synchronisation Triggers}
Various events may trigger the cluster-wide state synchronisation. Based on the locality of an event, we distinguish external and local synchronisation triggers in SDN controller.

\textbf{Locally activated Triggers.}
\begin{itemize}
\item Following the exhaustion of an execution credit set $C^{{Op}}_{K_I}$, the controller $K_I$ must contact other controllers from cluster set $K_N$ to lease additional execution credits, whereby $K_I \notin K_N$.
 \item Following the exhaustion of a resource credit set $C^{S}_{K_I}$ the controller $K_I$ must contact other controllers from cluster set $K_N$, where $K_I \notin K_N$ to lease additional resource credits.
\item If maximum duration of non-synchronisation period  $T^{S}_{max}$ is exceeded, the local controller propagates its current state version to all other cluster participants and resets the credit set $C^{S}_{K_I}$ / $C^{{Op}}_{K_I}$.
\item On unsuccessful commit (transaction abort) of an operation or identified divergent states, the resource owner assumes a conflict has happened. It retrieves the current state of other cluster participants to identify and resolve the local merge conflict.
\end{itemize}

\textbf{Externally activated Triggers.}
\begin{itemize}
\item Following an exhaustion of the execution or resource credit set in one controller in cluster, another controller replica is triggered to synchronise its state  and lease additional resource or execution credits.
\item If maximum duration of non-synchronisation period $T^{S}_{max}$ is exceeded at one controller replica, the replica initiates the cluster-wide synchronisation and all other controllers are triggered to accept the update.

\end{itemize}

\subsection{Algorithm}

Algorithm \ref{alg1} depicts the state synchronisation procedure for state $S_A$ and adaptation of active consistency level $CL_{S_A}$. On observed state $S_A$ modification in a remote controller $K_{rem}$, controller $K_{loc}$ proceeds to adapt the  consistency level $CL_{S_A}$ for state $S_A$ based on locally identified conflict-resolve  and result suboptimality costs ($C_{cflct}$ and $C_{sbptml}$, respectively) and given reference CL threshold maps $max_{S_A}[CL_{S_A}^{curr}]$ and $min_{S_A}[CL_{S_A}^{curr}]$. For simplicity, cost thresholds are manually specified by the SDN application which operates on the state. 

\begin{algorithm}
	\SetAlgoLined
	\DontPrintSemicolon
	
	\SetKwInOut{KwData}{Input}
	\SetKwInOut{KwResult}{Output}
	\SetKwInOut{KwInit}{Initialization}
	\SetKwProg{Fn}{Function}{}{}
	\SetKwProg{KwUpon}{upon}{}{}
	
	\KwData{\;			\PrintSemicolon
			State-update $S_A^{K_{rem},V_{rem}}$ with version vector $V_{rem}$ originated at remote controller $K_{rem}$\;
			Buffer of outstanding state-updates \emph{stateUpdateQueue}\;
			Initial active consistency level $CL_{S_A}$\;
			Mapping of maximum non-synchronisation durations $mapSyncPeriod[\ ]$ to various consistency levels\;
			Mapping of synchronisation credit set sizes $mapResourceCredits[\ ]$ to various consistency levels\;
			Mapping of minimum and maximum cost tresholds $min_{S_A}\lbrack\ \rbrack$ and $max_{S_A}[\ ]$ to various consistency levels\;
			}
			
	\KwInit{Number of iterations n = 0\;}
	\KwResult{Adapted consistency level $CL_{S_A}^{new}$\;}	
	
	\BlankLine
	
			\KwUpon{updated(stateUpdateQueue):} {
			
			\BlankLine
						
			$i = ++n$\;
			{$S_A^{K_{rem},V_{rem}} \leftarrow stateUpdateQueue.pop()$\;
			($S_{Anew}^{K_{loc},V_{loc}}$) $\leftarrow$ merge($S_A^{K_{loc},V_{loc}}$, $S_A^{K_{rem},V_{rem}}$)\;
			
			\BlankLine
			
			\If{conflictDetected($S_A^{K_{loc},V_{loc}}$) = true}
			{$C_{cflct} \leftarrow $ handleConflict($S_{Anew}^{K_{loc},V_{loc}}$)\;}
			
			\BlankLine
			
			\If{subOptimalityDetected($S_{Anew}^{K_{loc},V_{loc}}$) = true}
				{$C_{sbptml} \leftarrow $ handleSuboptimality($S_{Anew}^{K_{loc},V_{loc}}$)\;
				\BlankLine
				}
			
			\BlankLine
			
			$C_{S_A,sum}^i \leftarrow C_{cflct}+C_{sbptml}$\;
			
			\BlankLine
			
			$C^{S_A}_{accum} = \sum_{0}^{n} C_{S_A,sum}^i$\;
			$CL_{S_A}^{new} = adaptCL(C^{S_A}_{accum}, CL_{S_A})$\;
			
			\BlankLine
			
			\If{$CL_{S_A}^{new} \ne CL_{S_A}$}{
				$T^{S_A}_{max} = mapSyncPeriod[CL_{S_A}^{new}]$\;
				
				$C^{S_A}_{K_{loc}}  = mapResourceCredits[CL_{S_A}^{new}]$\;
				
				//$C^{{Op}}_{K_{loc}} = ma
				pExecutionCredits[CL_{S_A}^{new}]$\;
				
				genClusterEvent($CL_{S_A}^{new}, CL\_MOD$)\; 
				$CL_{S_A} = CL_{S_A}^{new}$
				}		
			
			\BlankLine
						
			localCommit($S_{Anew}^{K_{loc},V_{loc}}$)\;
		}
		
	\BlankLine
				
	\If{$C_{S_A,sum}.length() > observationWindowSize$}
	{$C_{S_A,sum}.shift(1)$\;
	 $n--$\;}
}
			\BlankLine
			
			\KwUpon{elapsed($Timer(T^{S_A}_{max})$):}{
				\If{$S_A^{K_{loc},V_{loc}}.updated()$}
				{genClusterEvent($S_A^{K_{loc},V_{loc}}, SYNC$)\;}
				
				\BlankLine
				
				$Timer(T^{S_A}_{max}).reset()$
			}
			
			\BlankLine
			
			\Fn{adaptCL ($measuredCost$, $CL_{S_A}^{curr}$)}{
				\uIf{$measuredCost>max_{S_A}[CL_{S_A}^{curr}]$}
				{return $CL_{curr}.tighten()$\;}
				\uElseIf{$measuredCost<min_{S_A}[CL_{S_A}^{curr}]$}
				{return $CL_{curr}.relax()$\;}
				\uElse
				{return $CL_{curr}$}
			}
\caption{State synchronisation procedure and adaptation of active state consistency level $CL_{S_A}$ for state $S_A$ in an SDN controller}
\label{alg1}
\end{algorithm}

In Algorithm \ref{alg1}, controller $K_{rem}$ triggers a state synchronisation event by sending a state update to controller $K_{loc}$. $K_{loc}$ then initiates the local state synchronisation as follows:
\begin{enumerate}
	\item Lines 3-4: Any obvious state-conflicts are detected by controller $K_{loc}$ (e.g. by using version vector \cite{Jank2000} comparison to determine state-update causality).
	\item Lines 5-7: In case of an identified version conflict, state conflict cost $C_{cflct}$ is computed based on cost of conflict-resolve strategy utilized to converge the views on state $S_A$.
	\item Lines 8-10: The SDN application computes the induced suboptimality of local result in previous non-synchronisation period and outputs the cost of suboptimality $C_{sbptml}$. Computation of this cost is specific to application logic. We show how an exemplary routing application can provide this implementation in Section \ref{simulation}.
	\item Line 11-13: Based on frequency and amplitude of $C_{cflct}$ and  $C_{sbptml}$, the active CL is adapted. The depicted approach assumes a simple threshold-based assignment of CLs, where the minimum and maximum thresholds associated with a CL are pre-defined.
	\item Lines 14-20: A new timer duration of non-synchronisation period is set. Depending on granularity of consistency design, either resource synchronisation credit set 	$C^{S_A}_{K_{loc}}$ or execution credit set $C^{{Op}}_{K_{loc}}$ (see Subsection \ref{sec:model}) for isolated modifications of state $S_A$ in $K_{loc}$ is loaded according to newly assigned CL. CL update is then propagated to all controllers that keep track of state $S_A$.
\end{enumerate}

After the state synchronisation procedure has determined the new consistency level $CL_{S_A}^{new}$, controllers that hold state $S_A$ modify their timers $T^{S_A}_{max}$ accordingly. If execution credits are assigned for operations that manipulate $S_A$, the number of execution credits allocated for isolated executions of these operations is lowered or raised according to $CL_{S_A}^{new}$. Analogously, if resource credits are assigned for modifications of state $S_A$, the number of resource credits allocated for isolated modifications is lowered or raised according to $CL_{S_A}^{new}$.

\subsection{Conflict Detection and Remediation}

Our eventually consistent SDN control plane assumes isolated state updates initiated by different controller instances at arbitrary points in time. Hence, controllers may operate on diverging isolated views of a shared state. Eventual propagation of one replica's state may lead to state conflicts in other replicas. In the case of network partitions or node failures,  the minority of nodes may lag behind the majority. When network partitions are merged or nodes recover from failure, the rejoined controller replicas might hold diverged states. A conflict handling strategy in such scenarios can be described in three dimensions:

\begin{enumerate}
	\item Identification of state-merge conflicts
	\item Resolution of state-merge conflicts
	\item Proactive avoidance of state-merge conflicts
\end{enumerate}

 While the method referenced in Algorithm \ref{alg1} is based on version vector comparison, advanced Distributed Database Management Systems (DDBMS) implement various strategies to identify synchronisation conflicts: 
\begin{itemize}
	\item \emph{Read-repair} \cite{Lakshman}: When a client query is made, DDBMS performs a digest query against all replicas and pushes the most recent version to out-of-date replicas.
	\item \emph{Anti-entropy}: Controllers periodically pool other replica's state and decide if their state views differ. Riak KV\footnote[1]{Riak KV - http://basho.com/products/riak-kv/}, a distributed NoSQL database, relies on a Merkle tree implementation for efficient divergent state discovery. Replicas in Riak recursively compare their trees until the conflict is localized. Similarly, ONOS \cite{Berde:2014:OTO:2620728.2620744} randomly picks up another replicated instance at 3-5 second intervals, and synchronises the respective topology views.
	\item Causality inference using \emph{version vectors}, version clocks, wall-clock time or similar update-version tracking mechanisms \cite{Lu2008, Jank2000}. By comparing logical clocks exchanged between actors on a state, concurrent quorum writes can be identified. 
\end{itemize}

Arising conflicts may lead to invalidation of some system invariants. In our model, each conflict is associated with a resolution cost $C_{cflct}$. The conflict resolution cost depends on the conflict weight, which is an application criteria, and the induced penalty time of conflict-repair, which is a property of the conflict resolve strategy. Different conflict resolve mechanisms can be considered with their respective cost:

\begin{itemize}
\item \emph{Update invalidation}: Operations are roll-backed and a reference consistent state is determined and applied at all conflicting replicas.
\item \emph{Replica-ID/Priority-based}: Specific replicas are preferred over others when comparing the state vectors.
\item \emph{State convergence on all replicas using commutative operations}: System makes assumption that all operations on target state commute  \cite{shapiro2011conflict}, or transforms the non-commutative to commutative operations \cite{li2012making}. 
\item \emph{Last-writer wins}: If causality of operation is deducible (e.g. by comparing the wall-clock), system selects the latest update. This strategy often leads to loss of one of the concurrent updates \cite{lloyd2011don}.
\item \emph{Manual Selection}: Client is offered conflicting state versions and must to select a single preferred version.
\end{itemize}


%% file: evaluation.tex
\subsection{Simulation of Concurrent Path Computations and Resource Reservations with Multiple SDN Controllers}
\label{simulation}
To evaluate the effect of consistency level adaptation, we have developed a cluster-aware path-finding SDN controller application which measures and logs the suboptimality of routing results, where suboptimality is a function of \emph{execution credit set} size, number of controller replicas, active consistency level, traffic model and network topology size. The application utilizes a bandwidth-constrained Dijkstra implementation to identify and reserve paths for uniformly selected source and destination pairs in a variable-size grid network. 

As the underlying network design and traffic patterns may bias the results experienced in practice, various topology sizes and traffic models were evaluated. The grid topology size was scaled between 5x5 to 25x25 vertices, with directed edges modeled as 1GbE links. The traffic models consider uniform specification of a flow bandwidth requirements in [1,30] Mbps range. We include an access control mechanism for new flow requests, which ensures that the edges whose 1GbE bandwidth capacities are exceeded during the reservation procedure are pruned before the Dijkstra routing procedure execution can take place. For simplification, the cost function which provides the cost input $C_E$ for an edge $E$, considers the total sum of flows configured on that edge:

\begin{equation}
C_E = \sum^{\#flows_E}_{i=0}1
\label{costfunction}
\end{equation}
 
Following an execution of a path finding algorithm in operation $add-flow$, bandwidth resources are reserved at the edges of the computed path. Whenever bandwidth utilization on an edge exceeds 80\%, for every new flow, an existing, uniformly and randomly selected flow is removed, hence allowing for embedding of a large number of sequential flow requests and realistic results. 

To introduce concurrency in execution, multiple  SDN Controller instances execute the path finding algorithm $add-flow$ in isolation from other instances. Hence  their local values of $C_E$ might differ during the non-synchronisation period. The synchronisation trigger fires after an assigned set of execution credits $C_{K_N}^{{Op}}$ is exceeded on every controller instance $K_N$. The controllers then synchronise the costs of edges $C_E$ and converge to the same state. We vary the execution credits $C_{K_N}^{{add-flow}}$, allocated for the path-finding operation \emph{add-flow} and focus on identifying the trade-off between the frequency of cluster-wide synchronisations and the result suboptimality which is formally defined as:

\begin{equation}
	D_{subopt} = \frac{O_{optimal}} {O_{measured}} 
\end{equation}
 
where $O_{optimal}$ is the cost of \emph{true} optimal path (computed as if all reservation updates in system were \emph{strictly serialized}); and $O_{measured}$ is the actual measured cost of a sampled path identified by an isolated instance during the non-synchronisation period. As each path computation in a controller instance $K_N$ only considers the reservation updates made during the non-synchronisation period on the local executor instance, in terms of total path cost, non-optimal paths with $D_{subopt} \ne 1$ could be determined. In our simulation, isolated reservations made by different instances of SDN controllers have often lead to concurrent reservation of bandwidth resources on the same edge, hence sacrificing the optimality of cheapest path finding. The suboptimality $D_{subopt}$ is computed after the synchronisation trigger fires, followed by a cluster-wide synchronisation of $C_E$.  

In the deployed eventually consistent model, reservation state modifications are instantaneous. While the cluster-wide synchronisation at the end of a non-synchronisation period is implemented as a blocking task, all intermediate local state changes are instantaneous updates, hence providing obvious response time benefits compared to a strong consistent approach. Consistency levels assigned for a state are defined by the amount of execution credits $C_{K_N}^{add-flow}$ assigned to cluster participants for isolated execution of operation \emph{add-flow}. 

\subsection{Experimental results}

To evaluate the effects of different deployed topologies and traffic models on experienced routing suboptimality, we vary the consistency level and hence the number of isolated executions  $C_{K_N}^{add-flow}$ per controller $K_N$. Our static mapping of amount of isolated executions of $add-flow$ operations to consistency level $CL_i$ is shown in Table \ref{tab:mapTable}. By manipulating the active consistency level, a controller instance $K_N$ executes a lowered or raised number of path finding executions in isolation from other instances $K_I \ne K_N$.

\begin{table} 
	\centering
	\begin{TAB}(r,1cm,0.5cm)[1.25pt]{|c|c|c|c|c|c|c|c|c|c|c|c|}{|c|c|}
		$\mathbf{Consist.Lvl}$ & $CL_1$ & $CL_2$ & $CL_3$ & $CL_4$ & $CL_5$ & $CL_6$& $CL_7$ & $CL_8$ & $CL_9$ & $CL_{10}$ & $CL_{11}$ \\
		$\mathbf{C^{add-flow}_{K_N}}$ & 2 & 3 & 5 & 9 & 17 & 25 & 33 & 41 & 49 & 57 & 65\\
	\end{TAB}

	\caption{Static mapping of the consistency level $CL_N$ to number of path finding execution credits $C^{add-flow}_{K_N}$ for operation \emph{add-flow} that may run in isolation.}
	\label{tab:mapTable}
\end{table}

\begin{figure} [H]
	\centering
	\includegraphics[scale=.50]{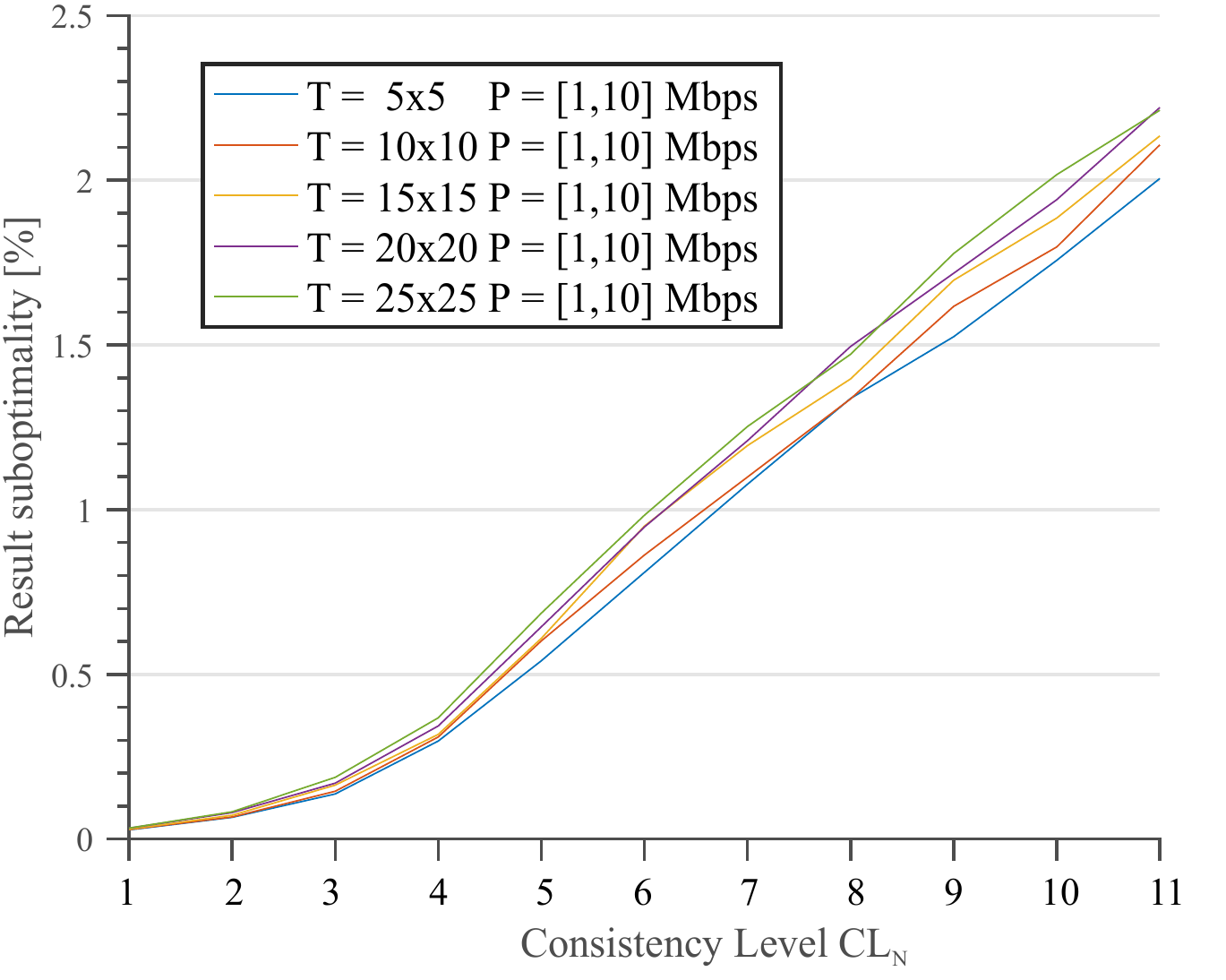}
	\caption{Measured Dijkstra routing inefficiency for variable topology sizes $T$, variable consistency levels $CL_N$ (as per Table \ref{tab:mapTable}), and traffic flows with uniformly distributed bandwidth requirement $P=[1,10]$ Mbps.  The cost suboptimality scales with the strictness of active consistency level and topology size. Our cost function, described in Equation \ref{costfunction}, considers the sum of flows placed on edges as edge cost, and bandwidth capacity as admission constraint. The curves may differ slightly for cost functions that consider other cost inputs (e.g. reserved bandwidth or buffer size).
	}
	\label{fig:routIneffTopo}
\end{figure}

According to Figure \ref{fig:routIneffTopo}, compared to smaller topology sizes, large topologies lead to higher result suboptimality $D_{subopt}$. A possible explanation for this phenomenon is that with the implemented \emph{bounded execution credit set} model, state synchronisation is triggered after an execution credit set is exceeded. In the case of large topologies, a single \emph{add-flow} execution can modify a higher number of switches on a computed path than in the case of smaller topologies. Hence consistency management using \emph{resource credit sets} might possibly lead to lower and uniform suboptimality for variable topology sizes. Furthermore, as expected, the $D_{subopt}$ of a routing operation scales linearly with the duration of non-synchronisation period during which the routing operations are executed in isolation. For a setup of $N=3$ controllers, each executing 65 concurrent flow additions in worst case, compared to a strongly consistent setup, 99th percentile of result suboptimality for the eventually consistent setup peaks at 2.22\%, a fairly low inefficiency.

\begin{figure} [H]
	\centering
	\includegraphics[scale=.50]{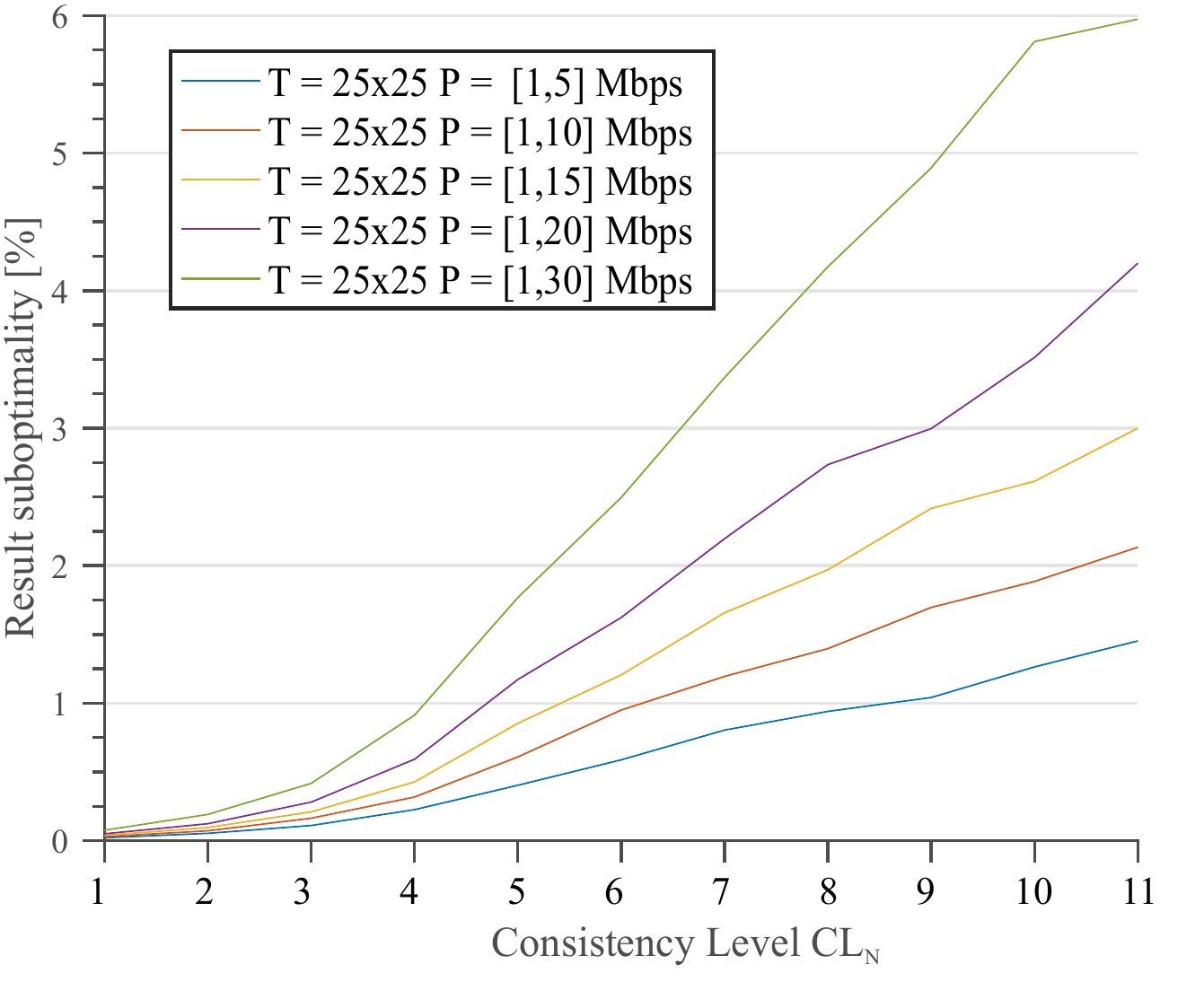}
	\caption{Measured Dijkstra routing inefficiency for variable traffic models  $P$, variable consistency levels $CL_N$ (as per Table \ref{tab:mapTable}), and a static 25x25 nodes grid topology. The cost suboptimality scales with the strictness of consistency levels and traffic models. Compared to micro-flows, the elephant-flows are more often evaluated as cost-inefficient paths, as larger amounts of resource are reserved on every flow admission and the 1Gbps edge capacity invariant is invalidated more frequently.  }
	\label{fig:routIneffModel}
\end{figure}

With regards to correlation between consistency level and result suboptimality, Figure \ref{fig:routIneffModel} depicts similar behaviour. It also shows how variation in traffic patterns influences performance of an eventually consistent system. By intelligent variation of assignment of execution credit amount and consistency level, bounding of  experienced suboptimality to an arbitrary target value is possible, regardless of active traffic patterns.

\begin{figure} [H]
	\centering
	\includegraphics[scale=.50]{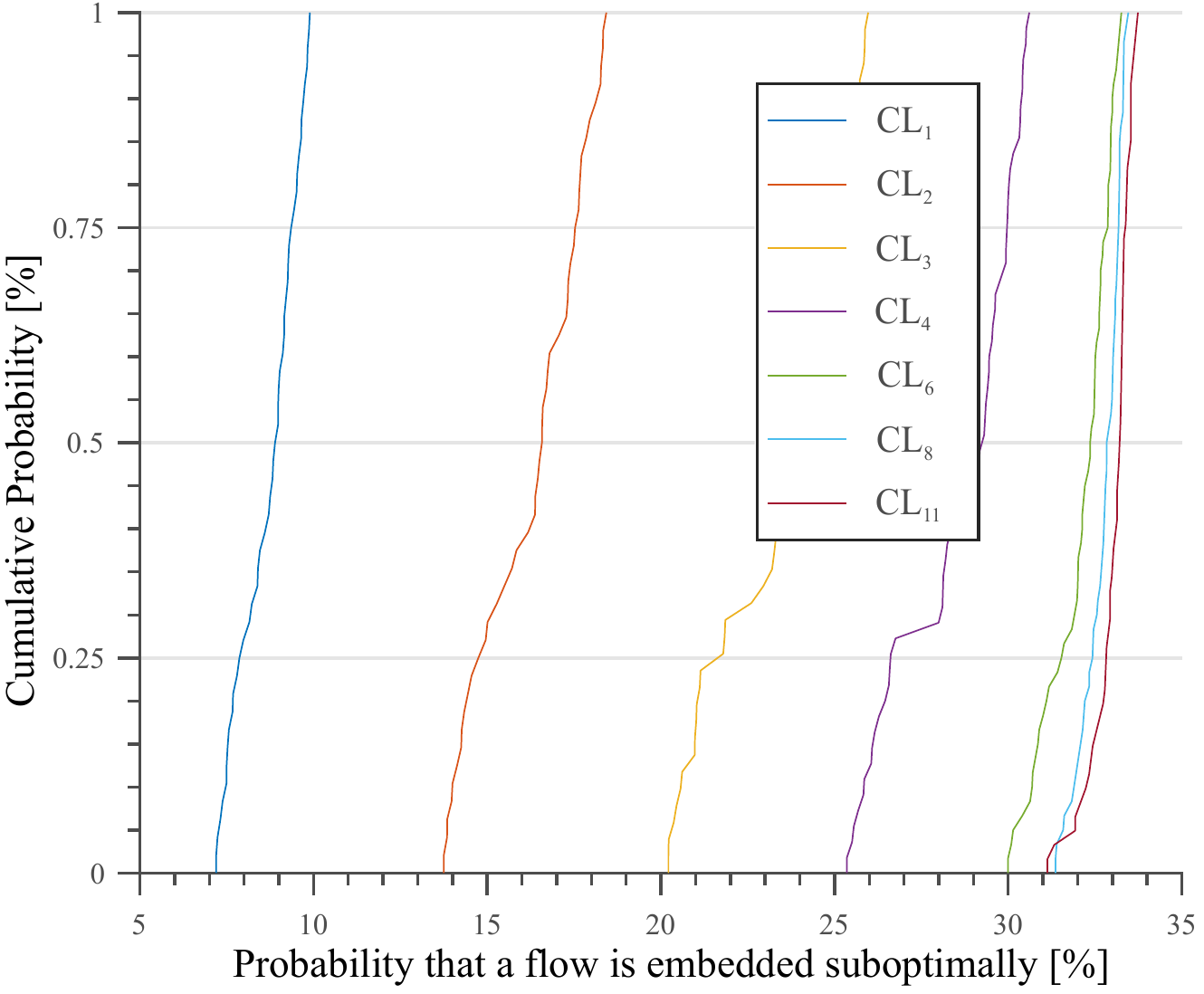}
	\caption{CDF of probability that during the non-synchronisation period a flow is embedded suboptimally. Cumulative probability is determined over all possible combinations of traffic models and topology sizes. With relaxation of the applied consistency level $CL_N$ (as per Table \ref{tab:mapTable}), probability rises that an identified reference path is suboptimal at end of its non-synchronisation period. With more relaxed consistency levels $CL_N > CL_6$, as many as 34\% of computed paths were identified as suboptimal at the end of non-synchronisation period. Regardless of this probability, Figures \ref{fig:routIneffTopo} and \ref{fig:routIneffModel} show that the qualitative inefficiency of suboptimal path cost always stays low.}
	\label{fig:pathChangeProb}
\end{figure}

Figure \ref{fig:pathChangeProb} portrays the cumulative probability of any single flow request arriving at the beginning of non-synchronisation period, for which initially an optimal solution is identified; being flagged as a suboptimal path at the time of cluster-wide synchronisation (at the end of non-synchronisation period). However, even when the probability of finding a suboptimal path is as high as 35\% in worst case, the overall suboptimality of identified path is acceptable, as shown in Figures \ref{fig:routIneffTopo} and \ref{fig:routIneffModel}.

\begin{figure} [H]
	\centering
	\includegraphics[scale=.50]{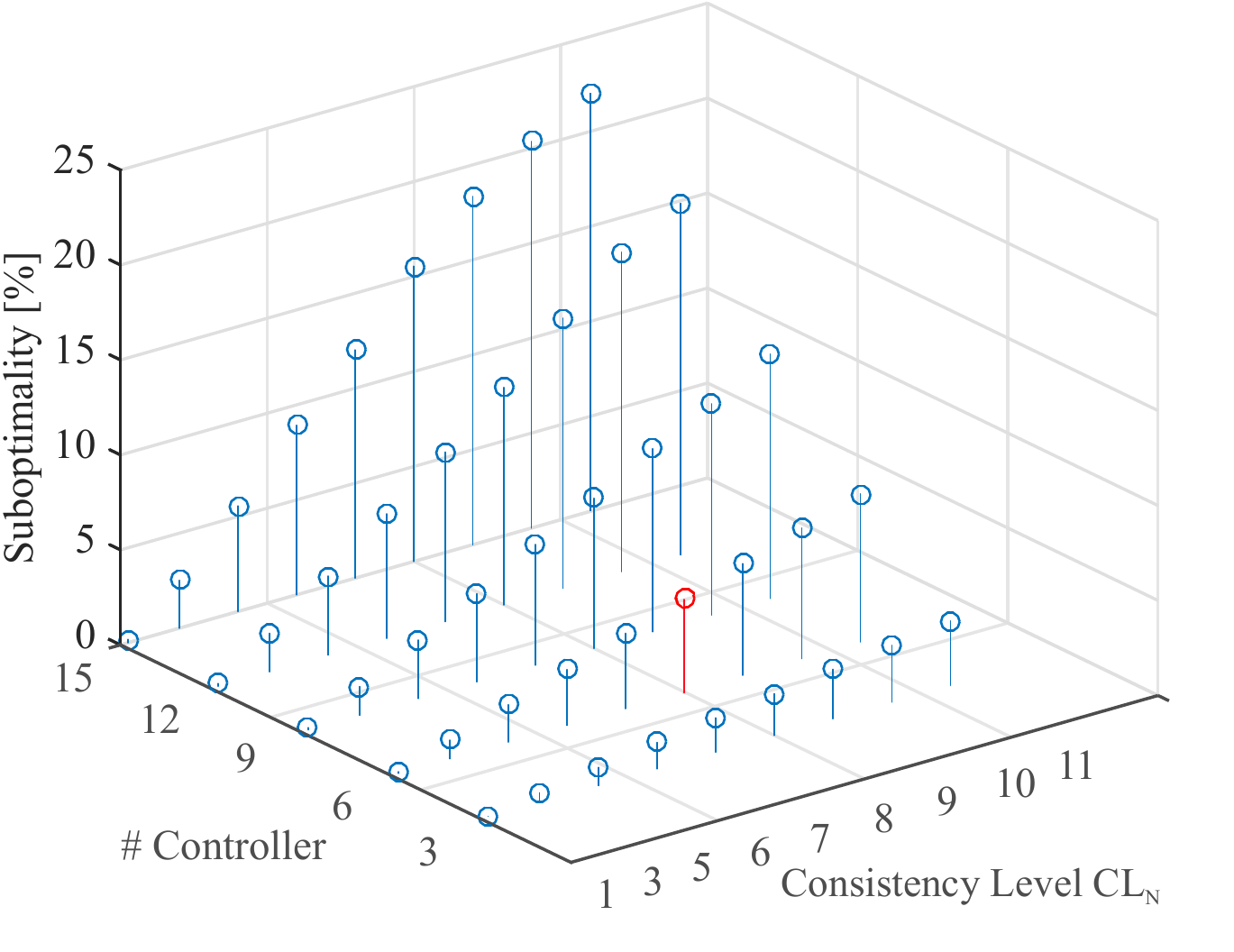}
	\caption{Load-balancing of path requests over a variable-size cluster of SDN controllers. The average suboptimality of sampled flows is determined over all possible combinations of traffic models and topology sizes. For different $CL_N$, each controller instance executes a variable number of \emph{add-flow} operations in isolation (as per Table \ref{tab:mapTable}). The state of admissioned flows on every edge $E$ (and hence the cost $C_E$) is distributed to all controller instances at the end of non-synchronisation period.}
	\label{fig:mapCluster}
\end{figure}

Concurrent execution of an operation allows for faster handling of a large set of same-type requests. Figure \ref{fig:mapCluster} depicts the measured mean suboptimality when handling a batch of 140k path requests for each combination of traffic model and topology size. We vary the size of cluster between 3 to 15 controller replicas. The consistency level defines the number of path finding executions executed in parallel at every of the available controller instances. The suboptimality is shown to scale with the number of concurrent path finding executions, and hence peaks at 25\% for the largest cluster size of 15 controllers, and the most-relaxed consistency level $CL_{11} \rightarrow C_{K_N}^{{add-flow}} = 65$ isolated operations per controller.

Figure \ref{fig:mapCluster} shows that a cluster of 6 SDN controllers is able to cope with consistency level set to $CL_8$ (41 isolated requests per instance), while limiting the path cost difference to less than 6\% compared to similar but strong consistent setup. In scenarios where flows are short-lived and higher cost inefficiencies can be tolerated, adaptation of the consistency level assigns higher shares of execution credits to controller instances.  A consistency level adaptation mechanism,  such as the threshold-based approach introduced in Algorithm \ref{alg1}, oscillates the experienced suboptimality around a named target value, while  minimizing the flow setup latencies independent of the network topology and traffic model at hand.


%% file: conclusion.tex
An eventually-consistent SDN DCP paves the way for scalable control plane designs. Our algorithm for adaptation of consistency levels leverages observed frequency and weight of conflicts in order to find a consistency level appropriate for targeted system optimality and correctness. In terms of response delay, enabling non-synchronised global switch configurations is especially efficient when working with short-lived flows that require fast response. By not relying on costly consensus after every single resource state update, end-clients in network can profit from shortened request-handling time in the SDN controller. If state synchronisation conflicts occur and correctness is endangered, our system adapts autonomously to a more appropriate consistency level. Threshold-based runtime modification of consistency levels considers costs of conflict in order to approximate the optimal trade-off between correctness and performance. We have shown by simulation that eventually consistent DCP can provide limited inefficiency compared to its strong consistent counterpart. 

Eventual consistency sacrifices some degree of correctness in order to provide performance. While failure of a controller may lead to delayed or lost events, achieving zero-loss property is a challenging research topic, even in a strong consistent DCP \cite{Katta:2015:RCF:2774993.2774996}.  Fault-tolerance properties of our consistency model were not considered in this paper but need further investigation. While strong and eventually consistent DCPs were compared in terms of cost-related metrics, time- and message-overhead metrics were left for later comparison. To this end, trade-offs between short execution time and blocking period duration incurred by conflict resolve procedure require further attention. Trade-offs between synchronisation overhead and result inefficiency when using either \emph{execution} or \emph{resource} credit sets are an additional open point for investigation. Finally, dynamic allocation of synchronisation credits based on reinforcement learning-approximated consistency levels could provide a more sophisticated alternative to the threshold-based approach presented in this paper.

%% file: IEEEPaper.bbl
\begin{thebibliography}{10}
\providecommand{\url}[1]{#1}
\csname url@samestyle\endcsname
\providecommand{\newblock}{\relax}
\providecommand{\bibinfo}[2]{#2}
\providecommand{\BIBentrySTDinterwordspacing}{\spaceskip=0pt\relax}
\providecommand{\BIBentryALTinterwordstretchfactor}{4}
\providecommand{\BIBentryALTinterwordspacing}{\spaceskip=\fontdimen2\font plus
\BIBentryALTinterwordstretchfactor\fontdimen3\font minus
  \fontdimen4\font\relax}
\providecommand{\BIBforeignlanguage}[2]{{%
\expandafter\ifx\csname l@#1\endcsname\relax
\typeout{** WARNING: IEEEtran.bst: No hyphenation pattern has been}%
\typeout{** loaded for the language `#1'. Using the pattern for}%
\typeout{** the default language instead.}%
\else
\language=\csname l@#1\endcsname
\fi
#2}}
\providecommand{\BIBdecl}{\relax}
\BIBdecl

\bibitem{Panda:2013:CN:2491185.2491186}
A.~Panda, C.~Scott, A.~Ghodsi, T.~Koponen, and S.~Shenker, ``{CAP for
  Networks},'' in \emph{Proceedings of the Second ACM SIGCOMM Workshop on Hot
  Topics in Software Defined Networking}.\hskip 1em plus 0.5em minus
  0.4em\relax ACM, 2013, pp. 91--96.

\bibitem{Bailis:2013:ECT:2447976.2447992}
P.~Bailis and A.~Ghodsi, ``{Eventual Consistency Today: Limitations,
  Extensions, and Beyond},'' \emph{Commun. ACM}, vol.~56, no.~5, 2013.

\bibitem{Demers:1987:EAR:41840.41841}
A.~Demers, D.~Greene, C.~Hauser, W.~Irish, J.~Larson, S.~Shenker, H.~Sturgis,
  D.~Swinehart, and D.~Terry, ``{Epidemic Algorithms for Replicated Database
  Maintenance},'' in \emph{Proceedings of the Sixth Annual ACM Symposium on
  Principles of Distributed Computing}.\hskip 1em plus 0.5em minus 0.4em\relax
  ACM, 1987.

\bibitem{Vogels:2009:EC:1435417.1435432}
W.~Vogels, ``{Eventually Consistent},'' \emph{Commun. ACM}, vol.~52, no.~1, pp.
  40--44, Jan. 2009.

\bibitem{barbard1992demarcation}
D.~Barbard and H.~Garcia-Molina, ``{The Demarcation Protocol: A technique for
  maintaining linear arithmetic constraints in distributed database systems},''
  in \emph{International Conference on Extending Database Technology}, 1992,
  pp. 373--388.

\bibitem{Medved}
J.~Medved, R.~Varga, A.~Tkacik, and K.~Gray,
  ``\BIBforeignlanguage{Undetermined}{{OpenDaylight: Towards a Model-Driven SDN
  Controller architecture}}.''

\bibitem{Berde:2014:OTO:2620728.2620744}
P.~Berde, M.~Gerola, J.~Hart, Y.~Higuchi, M.~Kobayashi, T.~Koide, B.~Lantz,
  B.~O'Connor, P.~Radoslavov, W.~Snow, and G.~Parulkar, ``{ONOS: Towards an
  Open, Distributed SDN OS},'' in \emph{Proceedings of the Third Workshop on
  Hot Topics in Software Defined Networking}, 2014.

\bibitem{paxos}
L.~Lamport, ``{Paxos Made Simple},'' \emph{ACM SIGACT News 32}, Dec. 2001.

\bibitem{dcn}
T.~Benson, A.~Akella, and D.~A. Maltz, ``{Network Traffic Characteristics of
  Data Centers in the Wild},'' in \emph{Proceedings of the 10th ACM SIGCOMM
  Conference on Internet Measurement}.\hskip 1em plus 0.5em minus 0.4em\relax
  ACM, 2010.

\bibitem{184040}
D.~Ongaro and J.~Ousterhout, ``{In Search of an Understandable Consensus
  Algorithm},'' in \emph{2014 USENIX Annual Technical Conference (USENIX ATC
  14)}.\hskip 1em plus 0.5em minus 0.4em\relax USENIX Association, Jun. 2014,
  pp. 305--319.

\bibitem{ho2016fast}
C.-C. Ho, K.~Wang, and Y.-H. Hsu, ``{A fast consensus algorithm for multiple
  controllers in software-defined networks},'' in \emph{18th International
  Conference on Advanced Communication Technology}, 2016.

\bibitem{du2014clock}
J.~Du, D.~Sciascia, S.~Elnikety, W.~Zwaenepoel, and F.~Pedone, ``{Clock-RSM:
  Low-latency inter-datacenter state machine replication using loosely
  synchronized physical clocks},'' in \emph{2014 44th Annual IEEE/IFIP
  International Conference on Dependable Systems and Networks}.\hskip 1em plus
  0.5em minus 0.4em\relax IEEE, 2014, pp. 343--354.

\bibitem{petropoulossoftware}
G.~Petropoulos, F.~Sardis, S.~Spirou, and T.~Mahmoodi, ``{Software-defined
  inter-networking: Enabling coordinated QoS control across the internet},'' in
  \emph{23rd International Conference on Telecommunications}, 2016.

\bibitem{bianco2016role}
A.~Bianco, P.~Giaccone, S.~D. Domenico, and T.~Zhang, ``{The Role of
  Inter-Controller Traffic for Placement of Distributed SDN Controllers},''
  \emph{CoRR}, vol. abs/1605.09268, 2016.

\bibitem{7517394}
J.~Guck, M.~Reisslein, and W.~Kellerer, ``{Function Split between
  Delay-Constrained Routing and Resource Allocation for Centrally Managed QoS
  in Industrial Networks},'' \emph{IEEE Transactions on Industrial
  Informatics}, no.~99, 2016.

\bibitem{5gic}
U.~o.~S. 5G~Innovation~Centre, ``{5G Whitepaper: The Flat Distributed Cloud
  (FDC) 5G Architecture Revolution},'' 2016.

\bibitem{Botelho:2013:FCF:2570448.2570470}
F.~A. Botelho, F.~M.~V. Ramos, D.~Kreutz, and A.~N. Bessani, ``{On the
  Feasibility of a Consistent and Fault-Tolerant Data Store for SDNs},'' in
  \emph{Proceedings of the 2013 Second European Workshop on Software Defined
  Networks}.\hskip 1em plus 0.5em minus 0.4em\relax IEEE Computer Society,
  2013, pp. 38--43.

\bibitem{Bessani2014StateMR}
A.~N. Bessani, J.~Sousa, and E.~A.~P. Alchieri, ``{State Machine Replication
  for the Masses with BFT-SMART},'' in \emph{DSN}, 2014.

\bibitem{tootoonchian2010hyperflow}
A.~Tootoonchian and Y.~Ganjali, ``{HyperFlow: A distributed control plane for
  OpenFlow},'' in \emph{{Proceedings of the 2010 internet network management
  conference on Research on enterprise networking}}, 2010.

\bibitem{stribling2009flexible}
J.~Stribling, Y.~Sovran, I.~Zhang, X.~Pretzer, J.~Li, M.~F. Kaashoek, and
  R.~Morris, ``{Flexible, Wide-Area Storage for Distributed Systems with
  WheelFS},'' in \emph{NSDI}, vol.~9, 2009, pp. 43--58.

\bibitem{DBLP:journals/corr/PhemiusBL13}
K.~Phemius, M.~Bouet, and J.~Leguay, ``{{DISCO:} Distributed Multi-domain {SDN}
  Controllers},'' \emph{CoRR}, vol. abs/1308.6138, 2013.

\bibitem{Koponen:2010:ODC:1924943.1924968}
T.~Koponen, M.~Casado, N.~Gude, J.~Stribling, L.~Poutievski, M.~Zhu,
  R.~Ramanathan, Y.~Iwata, H.~Inoue, T.~Hama, and S.~Shenker, ``{ONIX: A
  Distributed Control Platform for Large-scale Production Networks},'' in
  \emph{Proceedings of the 9th USENIX Conference on Operating Systems Design
  and Implementation}, 2010, pp. 351--364.

\bibitem{Levin:2012:LCS:2342441.2342443}
D.~Levin, A.~Wundsam, B.~Heller, N.~Handigol, and A.~Feldmann, ``{Logically
  Centralized?: State Distribution Trade-offs in Software Defined Networks},''
  in \emph{Proceedings of the First Workshop on Hot Topics in Software Defined
  Networks}.\hskip 1em plus 0.5em minus 0.4em\relax ACM, 2012, pp. 1--6.

\bibitem{Yu:2000:DEC:1251229.1251250}
H.~Yu and A.~Vahdat, ``{Design and Evaluation of a Continuous Consistency Model
  for Replicated Services},'' in \emph{Proceedings of the 4th Conference on
  Symposium on Operating System Design \& Implementation - Volume 4}.\hskip 1em
  plus 0.5em minus 0.4em\relax USENIX Association, 2000.

\bibitem{version1}
``{OpenFlow Switch Specification: Version 1.5.0 (Protocol Version 0x06)},''
  Open Networking Foundation, Dec. 2014.

\bibitem{Wang2016DynamicSC}
T.~Wang, F.~Liu, J.~Guo, and H.~Xu, ``{Dynamic SDN controller assignment in
  data center networks: Stable matching with transfers},'' in \emph{INFOCOM},
  2016.

\bibitem{mantas2016consistent}
A.~Mantas and F.~M.~V. Ramos, ``Consistent and fault-tolerant {SDN} with
  unmodified switches,'' \emph{CoRR}, vol. abs/1602.04211, 2016.

\bibitem{Jank2000}
A.~Jankunas, ``{Design and Evaluation of a Continuous Consistency Model for
  Replicated Services},'' vol.~45, no.~5, pp. 964--968, 2000.

\bibitem{Lakshman}
A.~Lakshman and P.~Malik, ``{Cassandra: a decentralized structured storage
  system},'' \emph{ACM SIGOPS Operating Systems Review}, vol.~44, no.~2, pp.
  35--40, 2010.

\bibitem{Lu2008}
Y.~Lu, Y.~Lu, and H.~Jiang, ``{Adaptive consistency guarantees for large-scale
  replicated services},'' \emph{Proceedings of the 2008 IEEE International
  Conference on Networking, Architecture, and Storage}, pp. 89--96, 2008.

\bibitem{shapiro2011conflict}
M.~Shapiro, N.~Pregui{\c{c}}a, C.~Baquero, and M.~Zawirski, ``{Conflict-free
  replicated data types},'' in \emph{Symposium on Self-Stabilizing Systems},
  2011.

\bibitem{li2012making}
C.~Li, D.~Porto, A.~Clement, J.~Gehrke, N.~Pregui{\c{c}}a, and R.~Rodrigues,
  ``Making geo-replicated systems fast as possible, consistent when
  necessary,'' in \emph{Proceedings of the 10th USENIX Symposium on Operating
  Systems Design and Implementation (OSDI 12)}, 2012, pp. 265--278.

\bibitem{lloyd2011don}
W.~Lloyd, M.~J. Freedman, M.~Kaminsky, and D.~G. Andersen, ``{Don't settle for
  eventual: scalable causal consistency for wide-area storage with COPS},'' in
  \emph{Proceedings of the Twenty-Third ACM Symposium on Operating Systems
  Principles}.\hskip 1em plus 0.5em minus 0.4em\relax ACM, 2011, pp. 401--416.

\bibitem{Katta:2015:RCF:2774993.2774996}
N.~Katta, H.~Zhang, M.~Freedman, and J.~Rexford, ``{Ravana: Controller
  Fault-tolerance in Software-defined Networking},'' in \emph{Proceedings of
  the 1st ACM SIGCOMM Symposium on Software Defined Networking Research}.\hskip
  1em plus 0.5em minus 0.4em\relax ACM, 2015, pp. 4:1--4:12.

\end{thebibliography}
